\documentclass[prd,twocolumn,showpacs, showkeys]{revtex4}
\begin{document}
\newcommand{\dfrac}[2]{\frac{\displaystyle #1}{\displaystyle #2}}
\baselineskip=12pt
\title{Real-time thermal Schwinger-Dyson equation for
quark self-energy in Landau gauge \footnote{The project
supported by the National Natural Science Foundation of China.}\\
}
\author{Zhou Bang-Rong}
\affiliation{College of Physical Sciences, Graduate School of the
Chinese Academy of Sciences, Beijing 100049, China}
\affiliation{
CCAST (World Laboratory), P.O.Box 8730, Beijing 100080, China }
\date{March 30, 2005}
\begin{abstract}
By means of a formal expression of the Cornwall-Jackiw-Tomboulis
effective potential for quark propagator at finite temperature and
finite quark chemical potential, we derive the real-time thermal
Schwinger-Dyson equation for quark propagator in Landau gauge.
Denote the inverse quark propagator by $A(p^2)\not\!{p}-B(p^2)$,
we argue that, when temperature $T$ is less than the given
infrared momentum cutoff $p_c$, $A(p^2)=1$ is a feasible
approximation and can be assumed in discussions of chiral symmetry
phase transition problem in QCD.
\end{abstract}
\pacs{11.10.Wx; 12.38.Lg; 11.30.Rd; 12.38.Aw} \keywords{Real-time
thermal QCD, CJT effective potential, Schwinger-Dyson equation,
quark propagator, Landau gauge}
\maketitle
\section{Introduction\label{Intro}}
The Schwinger-Dyson (SD) equation for quark propagator is one of
strong tools to research dynamical chiral symmetry breaking. A
great deal of work on it in the zero temperature case has been
made and a full demonstration of chiral symmetry breaking in zero
temperature Quantum Chromadynamics (QCD) has been given
\cite{kn:1,kn:2,kn:3,kn:4,kn:5,kn:6,kn:7,kn:8,kn:9,kn:10}.
However, the Schwinger-Dyson approach of chiral symmetry at finite
temperature has not been fully explored. Such research is of
important significance for understanding of the chiral phase
transition at high temperature and even the color superconducting
phase transition at low temperature and high density in QCD
\cite{kn:11}, since the SD approach is a non-perturbative one
which will give more reliable results.\\
\indent In  present work on the SD approach of chiral phase
transition, most are based on the imaginary-time formalism of
thermal field theory \cite{kn:12} and very few were seen in the
literature which are based on the real-time formalism. In fact, in
the real-time formalism, one can directly use some results in the
zero temperature QCD consistent with the renormalization group
(RG) analysis, thermalize them by thermal transformation matrices
and obtain the expected expressions at finite temperature and
chemical potential. In particular, the research in zero
temperature case has shown that in the Landau gauge, the
calculations can be simplified greatly \cite{kn:9}. It is expected
that at a finite temperature, this advantage will be maintained if
the Landau gauge is still
taken. \\
\indent The SD equation for quark propagator can be directly
derived from the extreme value condition of corresponding
Cornwall-Jackiw-Tomboulis (CJT) effective potential \cite{kn:13}.
In fact, if being limited to derive the SD equation, one needs
only a formal expression for the effective potential rather than
its a final explicit one. Conversely, the discussions based the SD
equation of the mass function in quark propagator will be quite
useful to simplifying the explicit calculation of corresponding
CJT effective potential for quark propagator and
this is also one of purposes of our research in this paper.\\
\indent The paper is arranged as follows. In Sect.\ref{eff} we
will give a formal expression of CJT effective potential for quark
propagator at finite temperature and finite quark chemical
potential in the real-time formalism of thermal field theory and
from which in Sect.\ref{SD}, we will derive in Landau gauge the SD
equation for quark self-energy in finite temperature and finite
chemical potential case. In Sect.\ref{Ap2} an important possible
approximation for the quark propagator will be explored and its
practical significance for chiral phase transition problem in QCD
will be indicated.
\section{The thermal CJT effective potential for quark
propagator\label{eff}}  It is well known that at temperature $T=0$
and the quark chemical potential $\mu=0$, the global flavor chiral
symmetries in QCD will be spontaneously broken owing to the vacuum
quark-antiquark condensates $\langle\bar{\psi}\psi\rangle\neq 0$
\cite{kn:8,kn:9,kn:14,kn:15}, this makes quarks acquire their
dynamical masses. For research on chiral symmetry restoring which
could occur at a finite $T$ and $\mu$, the thermal effective
potential for quark propagator is essential. We will derive a
formal expression of the effective potential first by writing the
CJT effective action $\Gamma[G,G^*]$ in QCD for the quark
propagator $G$ and its conjugate $G^*$ at finite $T$ and $\mu$ by
\begin{equation}
\Gamma_T[G,G^*]=\Gamma_1[G,G^*]+\Gamma_2[G,G^*],
\end{equation}
where $\Gamma_1[G,G^*]$ and $\Gamma_2[G,G^*]$ are respectively the
contribution from one-loop vacuum diagram and two and more loop
vacuum diagrams without quark self-energy correction, since $G$ in
Eq.(1) is the exact quark propagator. In the real-time formalism
of thermal field theory, we have
\begin{eqnarray}
\Gamma_1[G,G^*]&=&-i{\rm Tr}(\ln S_TG^{-1}_{T})^{11}\nonumber
\\&&-i{\rm Tr}( S_T^{-1}G_{T})^{11}+i{\rm Tr 1},
\end{eqnarray}
where $S_T$ and $G_T$ are $2\times2$ thermal matrix propagators,
the superscript "$11$" represents the $11$ component of the
corresponding matrix and the $\rm{Tr}$ is in functional sense.
Assuming translational invariance, then in momentum space, we have
\cite{kn:16}
\begin{equation}
S_T(p)=M_p\tilde{S}(p)M_p,\;\; G_T(p)=M_p\tilde{G}(p)M_p
\end{equation}
with the thermal transformation matrix $M_p$ defined by
$$
M_p=\left(%
\begin{array}{cc}
  \cos\theta_p & -e^{\beta\mu/2}\sin\theta_p \\
  e^{\beta\mu/2}\sin\theta_p  & \cos\theta_p \\
\end{array}%
\right)   ,\;\; \beta=1/T $$ \ $$
\sin^2\theta_p=\theta(p^0)\tilde{n}(p^0-\mu)+\theta(-p^0)\tilde{n}(-p^0+\mu),
$$
\begin{equation}
\tilde{n}(p^0-\mu)=1/[e^{\beta(p^0-\mu)}+1]
\end{equation}
and
\begin{equation}
\tilde{S}(p)=\left(
\begin{array}{cc}
  S(p)& 0 \\
  0 & S^*(p) \\
\end{array}
\right)    , \;\; \tilde{G}(p)=\left(
\begin{array}{cc}
  G(p) & 0 \\
  0 & G^*(p) \\
\end{array}
\right)
\end{equation}
where
\begin{equation}
S(p)=i/(\not\!{p}+i\varepsilon),
\;\;S^*(p)=-i/(\not\!{p}-i\varepsilon),\;\not\!{p}=\gamma^{\mu}p_{\mu}
\end{equation}
and
\begin{eqnarray}
G(p)&=&i/[A(p^2)\not\!{p}-B(p^2)+i\varepsilon],\nonumber \\
G^*(p)&=&-i/[A(p^2)\not\!{p}-B(p^2)-i\varepsilon].
\end{eqnarray}
$G(p)$ and $G^*(p)$ are the complete propagators of quarks with
dynamical mass, where $A(p^2)$ and $B(p^2)$ have been assumed to
be real functions. Thus in the momentum space,
\begin{eqnarray}
\Gamma_1[G,G^*]&=& -\Omega iN_fN_c\left\{\big\langle{\rm tr}[\ln
S_T(p)G_T^{-1}(p)]^{11}\big\rangle\right.\nonumber \\
&&\left.+\big\langle{\rm
tr}[S_T^{-1}G_T(p)]^{11}\big\rangle-\big\langle{\textrm tr
1}\big\rangle\right\}\nonumber \\
&\equiv&-\Omega V_1[G,G^*]
\end{eqnarray}
where $N_f$ and $N_c$ are respectively the number of flavor and
color of the quarks, $\Omega$ is the volume of space-time,
$\textrm{tr}$ only represents the trace of  spinor matrices and
the denotation $\langle\cdots\rangle$ means the integration $\int
d^4p/(2\pi)^4$. By means of Eqs.(3)-(6) and the relation
\cite{kn:17}
$$ \ln
S_T(p)G_T^{-1}(p)=M_p\ln[\tilde{S}(p)\tilde{G}^{-1}(p)]M^{-1}(p),
$$
we obtain from Eq.(8) the one-loop effective potential
$V_1[G,G^*]$ corresponding to $\Gamma_1[G,G^*]$
\begin{widetext}
\begin{eqnarray}
V_1[G,G^*]&=& iN_fN_c\left(
\big\langle\cos^2\theta_p\big\{\textrm{tr}\ln[S(p)G^{-1}(p)]+\textrm{tr}[S^{-1}(p)G(p)]\big\}
\big\rangle\right.\nonumber \\
  &&\left.+\big\langle\sin^2\theta_p\big\{\textrm{tr}\ln[S^*(p)G^{*-1}(p)]+
  \textrm{tr}[S^{*-1}(p)G^*(p)]\big\}\big\rangle
  -\big\langle\textrm{tr}1\big\rangle\right).
\end{eqnarray}
\indent In the real-time formalism of thermal field theory, the
interaction Lagrangian between the quark fields $\psi$ and the
gluon fields $A_{\mu}^{a} (a=1,\cdots,8)$ can be expressed by
$$
{\cal{L}}_i=-\sum_{a=1}^{8}\sum_{r=1,2}g_0(-1)^{r+1}
\bar{\psi}^{(r)}\gamma^{\mu}A_{\mu}^{a(r)}\psi^{(r)}\lambda^a(\underline{N}_c),
$$
where $r=1$ and $r=2$ correspond to physical and (thermal) ghost
fields respectively, $\lambda^a(\underline{N}_c)$ is the generator
matrix of the color gauge group $SU_c(3)$ in the representation
$\underline{N}_c$ of the quarks and $g_0$ is the bare gauge
coupling constant. Assume that only the two-loop vacuum diagrams
are included in $\Gamma_2[G,G^*]$, then we can write in the
momentum space
\begin{eqnarray}
  \Gamma_2[G,G^*] &=&-\frac{i}{2}\big\langle\big\langle g_0^2\textrm{Tr}{
  \gamma^{\mu}\lambda^a(\underline{N}_c)G_T^{11}(p)\gamma^{\nu}\lambda^b(\underline{N}_c)
  G_T^{11}(q)[D_{\mu\nu}^{\prime ab}(p-q)]_T^{11}}\big\rangle\big\rangle N_f\Omega\nonumber \\
   &&+\frac{i}{2} \big\langle\big\langle g_0^2\textrm{Tr}{
  \gamma^{\mu}\lambda^a(\underline{N}_c)G_T^{12}(p)\gamma^{\nu}\lambda^b(\underline{N}_c)
  G_T^{21}(q)[D_{\mu\nu}^{\prime ab}(p-q)]_T^{12}}\big\rangle\big\rangle
  N_f\Omega \nonumber \\
  &\equiv&-\Omega V_2[G,G^*]
\end{eqnarray}
\end{widetext}
where $[D_{\mu\nu}^{\prime ab}(p-q)]_T$ is the complete thermal
gluon matrix propagator, $\langle\langle\cdots\rangle\rangle$
represents the integrations $\int\int d^4pd^4q/(2\pi)^8$ and  Tr
is now only for the quark-dependent matrices (flavor, color, and
spinor etc.). After the renormalization procedure is implemented,
we can replace $g^2_0$ by the RG-invariant running gauge coupling
$g^2\big[(p-q)^2\big]$, and $\big[D_{\mu\nu}^{\prime
ab}(p-q)\big]_T$ by the tree diagram gluon propagator
$\big[D_{\mu\nu}^{ab}(p-q)\big]_T$, i.e. we will take the
approximation \cite{kn:8, kn:9}
\begin{equation}
 g^2_0 \big[D_{\mu\nu}^{\prime ab}(p-q)\big]_T\simeq g^2\big[(p-q)^2\big]
 \big[D_{\mu\nu}^{ab}(p-q)\big]_T.
\end{equation}
In Eq.(10) the complete thermal vertices $\Gamma^{\mu(r)}(p,q)
(r=1,2)$ have been replaced by the tree diagram vertex
$\gamma^{\mu}$. This is because when $A(p^2)$ and $B(p^2)$ in the
inverse quark propagator $G^{-1}(p)$ are real functions, the
Ward-Takahashi (WT) identity at finite $T$ and $\mu$ (if the ghost
effect in the fermion sector is neglected) essentially identical
to the one at $T=\mu=0$, thus in the Landau gauge we can make the
couplings between the complete vertices $\Gamma^{\mu(r)}(p,q)
(r=1,2)$ which submit to the WT identity and the thermal gluon
propagator are equivalent to the one between the tree diagram
vertex $\gamma^{\mu}$ and the thermal gluon propagator
\cite{kn:18}. The tree diagram thermal gluon matrix propagator
$[D_{\mu\nu}^{ab}(k)]_T$ can be expressed by \cite{kn:16}
\begin{equation}
\big[D_{\mu\nu}^{ab}(k)\big]_T=\big[D_{\mu\nu}(k)\big]_T\delta^{ab},\;\;
\big[D_{\mu\nu}(k)\big]_T=\bar{M}_k\tilde{D}_{\mu\nu}(k)\bar{M}_k
\end{equation}
with the thermal transformation matrix $\bar{M}_k$ defined by
\begin{eqnarray}
\bar{M}_k&=&\left(
\begin{array}{cc}
  \cosh\Theta_k & \sinh\Theta_k \\
  \sinh\Theta_k & \cosh\Theta_k \\
\end{array}%
\right),\;\;\sinh\Theta_k=\sqrt{n(k^0)},\nonumber \\&&
n(k^0)=\dfrac{1}{e^{\beta|k^0|}-1}
\end{eqnarray}
and
\begin{equation}
\tilde{D}_{\mu\nu}(k)=\left(
\begin{array}{cc}
  D_{\mu\nu}(k) & 0 \\
  0 & D^*_{\mu\nu}(k) \\
\end{array}
\right).
\end{equation}
In Landau gauge,
\begin{equation}
\left\{
\begin{array}{c}
  D_{\mu\nu}(k)  \\
  D^*_{\mu\nu}(k) \\
\end{array}\right. =\dfrac{\mp i}{k^2\pm i\varepsilon}
\left(g_{\mu\nu}-\dfrac{k_{\mu}k_{\nu}}{k^2\pm
i\varepsilon}\right).
\end{equation}
By means of Eqs.(3)-(5) and Eqs. (11)-(14), we may obtain from
Eq.(10) the two-loop effective potential $V_2[G,G^*]$
corresponding to $\Gamma_2[G,G^*]$, expressed by
\begin{widetext}
\begin{eqnarray}
 V_2[G,G^*]&=& \frac{i}{2}N_fN_cC_2(\underline{N}_c) \big\langle\big\langle g^2\big[(p-q)^2\big]
 \textrm{tr}\big[\gamma^{\mu}G^{11}_T(p)\gamma^{\nu}G^{11}_T(q)\big]
 \big[D_{\mu\nu}(p-q)\big]^{11}_T\big\rangle\big\rangle \nonumber\\
  && -\frac{i}{2}N_fN_cC_2(\underline{N}_c)\big\langle\big\langle g^2\big[(p-q)^2\big]\textrm{tr}
 \big[\gamma^{\mu}G^{12}_T(p)\gamma^{\nu}G^{21}_T(q)\big]
 \big[D_{\mu\nu}(p-q)\big]^{12}_T\big\rangle\big\rangle.
\end{eqnarray}
with
\begin{equation}
G^{11}_T(p)=\cos^2\theta_pG(p)-\sin^2\theta_pG^*(p),\;\;
G^{12}_T(p)=-\cos\theta_p\;\sin\theta_p\;
e^{\beta\mu/2}[G(p)+G^*(p)]=-e^{\beta\mu}G_T^{21}(p),
\end{equation}
\begin{equation}
\big[D_{\mu\nu}(p-q)\big]^{11}_T=D_{\mu\nu}(p-q)-g_{\mu\nu}2\pi
n(p^0-q^0)\delta\big[(p-q)^2\big],\;\;\big[D_{\mu\nu}(p-q)\big]^{12}_T
=-g_{\mu\nu}e^{\beta|p^0-q^0|/2}2\pi
n(p^0-q^0)\delta\big[(p-q)^2\big]
\end{equation}
and
$$
N_cC_2(\underline{N}_c)=\sum_{a,b}\textrm{tr}\big[\lambda^a(\underline{N}_c)
\lambda^b(\underline{N}_c)\big],
\;\;C_2(\underline{N}_c)=\dfrac{N_c^2-1}{2N_c}.
$$
The total effective potential will be
\begin{equation}
V[G,G^*]=V_1[G,G^*]+V_2[G,G^*]
\end{equation}
\end{widetext}
\section{The thermal Schwinger-Dyson equation for quark
self-energy}\label{SD}

The effective potential $V[G,G^*]$ is in fact the functional of
the thermal propagator $\big[G(p)\big]_T^{rs}$ and
$\big[D_{\mu\nu}^{ab}(p-q)\big]_T^{rs} \; (rs=11,12)$. The
Schwinger-Dyson equation for quark self-energy will come from the
extremum condition
\begin{equation}
\dfrac{\delta V[G,G^*]}{\delta G_T^{11}(p)}=0.
\end{equation}
In view of Eq.(17),  equation (20) is equivalent to
$$
\left[\dfrac{\delta}{\delta G(p)}-\dfrac{\delta}{\delta
G^*(p)}\right]V[G,G^*]=0
$$
which, by using Eqs.(9), (16) and (19), can be reduced to
\begin{eqnarray}
\Sigma(p)&\equiv& [1-A(p^2)]\not\!{p}+B(p^2) \nonumber\\
   &=& -iC_2(\underline{N}_c)\big\langle g^2[(p-q)^2]\gamma^{\mu}G^{11}_T(q)
   \gamma^{\nu}[D_{\mu\nu}(p-q)]_T^{11}\big\rangle\nonumber\\
\end{eqnarray}  
Equation (21) is just the SD equation of the complete self-energy
$\Sigma(p)$ for single flavor and single color quark to one-loop
order at finite $T$ and $\mu$. It is noted that the second term in
the right-handed side of Eq.(16) has no contribution to Eq.(21).
By acting  $\textrm{tr}\!\!\not\!{p}$  and $\textrm{tr}$ on the
two sides of Eq.(21) separately, we can obtain the equations of
$A(p^2)$ and $B(p^2)$,
\begin{widetext}
\begin{equation}
A(p^2)=1+i\frac{C_2(\underline{N}_c)}{4p^2}\Big\langle
g^2\big[(p-q)^2\big]\textrm{tr}\big[\!\not\!{p}\gamma^{\mu}G^{11}_T(q)\gamma^{\nu}\big]
\big[D_{\mu\nu}(p-q)\big]^{11}_T\Big\rangle
\end{equation}
\begin{equation}
B(p^2)=-i\frac{C_2(\underline{N}_c)}{4}\Big\langle
g^2\big[(p-q)^2\big]\textrm{tr}\big[\gamma^{\mu}G^{11}_T(q)\gamma^{\nu}\big]
\big[D_{\mu\nu}(p-q)\big]^{11}_T\Big\rangle.
\end{equation}
From Eqs.(7) and (17) we have
\begin{equation}
G^{11}_T(q)=\big[A(q^2)\!\not\!{q}+B(q^2)\big]\left\{\dfrac{i}{A^2(q^2)q^2-B^2(q^2)+i\varepsilon}-
2\pi\sin^2\theta_q\delta\big[A^2(q^2)q^2-B^2(q^2)\big]\right\}.
\end{equation}
By means of Eqs.(18) and (24), and  the result that
$$
\textrm{tr}\big(\!\not\!{p}\gamma^{\mu}\!\not\!{q}\gamma^{\nu}\big)D_{\mu\nu}(p-q)=-i4E(p,q)
$$
with
\begin{equation}
E(p,q)=1-\frac{1}{2}\left[p^2+q^2+\frac{(p^2-q^2)^2}{(p-q)^2}\right]/(p-q)^2,
\end{equation}
equation (22) can be changed into
\begin{eqnarray}
 A(p^2)&=&1+i\frac{C_2(\underline{N}_c)}{p^2}\Big\langle
g^2\big[(p-q)^2\big]
A(q^2)\Big\{\frac{i}{A^2(q^2)q^2-B^2(q^2)+i\varepsilon}-
2\pi\sin^2\theta_q\delta\big[A^2(q^2)q^2-B^2(q^2)\big]\Big\} \nonumber\\
  && \times\Big\{-iE(p,q)+4\pi(p\cdot
  q)n(p^0-q^0)\delta\big[(p-q)^2\big]\Big\}\Big\rangle.
\end{eqnarray}
It is obtained that the integration in Eq.(26)
$$ \Big\langle g^2\big[(p-q)^2\big]
A(q^2)\frac{E(p,q)}{A^2(q^2)q^2-B^2(q^2)+i\varepsilon}\Big\rangle
=-i\int\frac{d^4\bar{q}}{(2\pi)^4}g^2[(\bar{p}-\bar{q})^2]
A(\bar{q}^2)\dfrac{E(\bar{p},\bar{q})}{A^2(\bar{q}^2)\bar{q}^2+B^2(\bar{q}^2)}=0,
$$
where we have made the changes of variables $\bar{p}^0=ip^0, \;
\bar{p}^i=p^i \;(i=1,2,3)$ and the same to $q\rightarrow \bar{q}$
after the Wick rotation, and the assumption that \cite{kn:9}
$$
g^2\big[(\bar{p}-\bar{q})^2\big]=\theta(\bar{p}^2-\bar{q}^2)g^2(\bar{p}^2)+
\theta(\bar{q}^2-\bar{p}^2)g^2(\bar{q}^2),
$$
and then consider the fact that
$$
\int d\Omega_{\bar{q}} E(\bar{p},\bar{q})=0.
$$
As a result, equation (26) is reduced to
\begin{eqnarray}
A(p^2)&=&1-2\pi\frac{C_2(\underline{N}_c)}{p^2}\Big\langle
g^2\big[(p-q)^2\big]A(q^2)
\Big\{E(p,q)\sin^2\theta_q\delta\big[A^2(q^2)q^2-B^2(q^2)\big]\nonumber\\
&&+\Big(\frac{1}{A^2(q^2)q^2-B^2(q^2)+i\varepsilon}+i
2\pi\sin^2\theta_q\delta\big[A^2(q^2)q^2-B^2(q^2)\big]
\Big)\cdot2(p\cdot
q)n(p^0-q^0)\delta\big[(p-q)^2\big]\Big\}\Big\rangle.
\end{eqnarray}
On the other hand, by using Eqs. (18) and (24),  equation (23) can
be changed into
\begin{eqnarray}
 B(p^2)&=&-i C_2(\underline{N}_c)\Big\langle g^2\big[(p-q)^2\big]B(q^2)\Big\{\frac{i}{A^2(q^2)q^2-B^2(q^2)+i\varepsilon}-
2\pi\sin^2\theta_q\delta\big[A^2(q^2)q^2-B^2(q^2)\big]\Big\}\nonumber \\
   && \times\Big\{\frac{-3i}{(p-q)^2+i\varepsilon}-8\pi n(p^0-q^0)\delta\big[(p-q)^2\big]
   \Big\}\Big\rangle.
\end{eqnarray}
Equations (27) and (28) are the required coupled equations of the
function $A(p^2)$ and $B(p^2)$ in the quark self-energy
$\Sigma(p)$ in Landau gauge in the real-time thermal QCD. It is
indicated that when $T=\mu=0$, all the thermal correction terms
can be removed, hence equations (27) and (28) will be reduced to
\begin{equation}
A(p^2)=1, \; \; B(p^2)=-i3C_2(\underline{N}_c)\Big\langle
g^2\big[(p-q)^2\big]
B(q^2)\frac{1}{q^2-B^2(q^2)+i\varepsilon}\cdot\frac{1}{(p-q)^2+i\varepsilon}\Big\rangle
\end{equation}
which are just the SD equations for quark self-energy function in
Landau gauge at $T=\mu=0$.
\end{widetext}
\section{Discussion of $A(p^2)$\label{Ap2}}
Equation (27) shows that, at a finite $T$ and $\mu$, $A(p^2)$ has
extra thermal correction terms to unity. Based on the
consideration of dimension, $A(p^2)$ can be generally written by $
A(p^2)=1-(T^2/p^2)F$, where $F$ depends on only the dimensionless
variants consisting of $T$, $p$, $B$,$\mu$ etc..  Assuming $F$ is
finite, then it can be deduced that at a high $T^2$ and a low
$p^2$, the thermal correction  terms would be not small and
$A(p^2)=1$ would not be a good approximation.  However, we will
argue that for some physically interesting range of $T$, the
thermal corrections of $A(p^2)$ are negligible hence the
approximation
$A(p^2)=1$ may still be valid. \\
\indent Denote the quark mass function by
\begin{equation}
m^2(q^2)=B^2(q^2)/A^2(q^2),
\end{equation}
then $A(p^2)$ can be expressed by
\begin{widetext}
\begin{eqnarray}
A(p^2)&=&1-\frac{C_2(\underline{N}_c)}{p^2}\int\frac{d^4q}{(2\pi)^3}\frac{1}{A(q^2)}
\Big\{g^2\big[(p-q)^2\big]E(p,q)\sin^2\theta_q\delta\big[q^2-m^2(q^2)\big]\nonumber\\
&&+g^2(0)\Big(\frac{1}{q^2-m^2(q^2)+i\varepsilon}+i
2\pi\sin^2\theta_q\delta\big[q^2-m^2(q^2)\big]
\Big)(p^2+q^2)n(p^0-q^0)\delta\big[(p-q)^2\big]\Big\}.
\end{eqnarray}
Let $q^2=m_1^2$ is a real root of the equation $q^2=m^2(q^2)$,
i.e. $m_1$ obeys the equation
\begin{equation}
 m_1^2=m^2(m_1^2),
\end{equation}
then we will have
\begin{equation}
\delta\big[q^2-m^2(q^2)\big]=\frac{1}{f(m_1^2)}\delta\big[q^2-m^2_1)\big],\;\;
f(m^2_1)=\Big|1-\frac{\partial m^2(q^2)}{\partial
q^2}\Big|_{q^2=m_1^2}.
\end{equation}
In fact, one can identify $m(q^2)$ with the running quark mass
function consistent with the renormalization group (RG) in QCD
\cite{kn:8,kn:9}, then it will be a descent function of $q^2$ and
we always have $f(m_1^2)\geq1$. To estimate the order of magnitude
of the thermal corrections of $A(p^2)$, we will take the
approximation
\begin{eqnarray*}
g^2\big[(p-q)^2\big]E(p,q)&\simeq&\theta(|p^2|-|q^2|)\;
g^2(p^2)\left\{1-\frac{1}{2}
\left[p^2+q^2+\frac{(p^2-q^2)^2}{p^2}\right]\frac{1}{p^2}\right\}\nonumber \\
 &&+\theta(|q^2|-|p^2|)\;g^2(q^2)\left\{1-\frac{1}{2}
\left[p^2+q^2+\frac{(p^2-q^2)^2}{q^2}\right]\frac{1}{q^2}\right\}
\end{eqnarray*}
and
$$
\frac{1}{A(q^2)}\cdot\frac{1}{q^2-m^2(q^2)+i\varepsilon}\simeq\frac{1}{A(m_1^2)f(m_1^2)}
\cdot\frac{1}{q^2-m_1^2+i\varepsilon}.
$$
Considering the results that
$$ \int d^4q
\sin^2\theta_q\delta(q^2-m_1^2)=4\pi T^2I_3(y_1,r)
$$
with
$$
 I_3(y_1,r)=\frac{1}{2}\int_0^{\infty}dx
 \frac{x^2}{\sqrt{x^2+y_1^2}}
 \left[\dfrac{1}{e^{\sqrt{x^2+y_1^2}-r}+1}+(-r\rightarrow r)\right],\;
  y_1=\frac{m_1}{T},\; r=\frac{\mu}{T}
$$
and
$$
\int d^4q n(q^0)\delta(q^2)=\frac{2}{3}\pi^3T^2,
$$
we can write
\begin{equation}
A(p^2)=1-a(p^2)/A(m_1^2),
\end{equation}
where
\begin{eqnarray}
 a(p^2)&=&\frac{1}{f(m_1^2)}\left\{
  \left[\frac{2}{3\pi^2}\tilde{g}^2(p^2)I_3(y_1,r)+\frac{g^2(0)}{9}\right]
 \frac{T^2}{(|p|^2+p_c^2)\varepsilon(p^2)}\right. \nonumber\\
   &&\left.+\frac{g^2(0)}{12\pi^2}\frac{p^2+m_1^2}{(|p|^2+p_c^2)\varepsilon(p^2)}\int_0^{\infty}\frac{dx}{e^x-1}
   \left[\ln\frac{p^2-m_1^2+2xT(p^0+|\vec{p}|)}{p^2-m_1^2+2xT(p^0-|\vec{p}|)}+
   (p^0\rightarrow -p^0)\right]\frac{T}{|\vec{p}|}\right\}
\end{eqnarray}
with
\begin{equation}
\tilde{g}^2(p^2)=\theta(m_1^2-|p^2|)g^2(m_1^2)\frac{p^2}{2(m_1^2+p^2_c)}\left(1-\frac{p^2}{m_1^2}\right)
+\theta(|p^2|-m_1^2)g^2(p^2)\frac{m_1^2}{2p^2}\left(1-\frac{m_1^2}{p^2}\right).
\end{equation}
\end{widetext}
In Eq.(35) we have introduced the infrared momentum cutoff $p_c^2$
by the replacement $1/p^2\rightarrow
1/(|p|^2+p_c^2)\varepsilon(p^2)$ with $\varepsilon(p^2)\equiv
p^2/|p^2|$ and in Eq.(36) by the replacement $1/2m_1^2 \rightarrow
1/2(m_1^2+p_c^2)$. When $p^2=m_1^2$, both $\tilde{g}^2(p^2)$ and
the integral with the logarithm functions in Eq.(35) are equal to
zeroes, hence we get
\begin{equation}
a(m_1^2)=\frac{g^2(0)T^2}{9f\big(m_1^2\big)\big(m_1^2+p_c^2\big)}.
\end{equation}
When $p^2\neq m_1^2$ and if $|p^2-m_1^2|>T^2$ is assumed, we can
expand the logarithm functions in Eq.(35) in power of
$2xT(p^0\pm|\vec{p}|)/(p^2-m_1^2)$ (noting that the main
contribution to the integral over $x$ comes from the region of
$x<1$), then in the approximation with the infrared momentum
cutoff $p_c^2$ included that
\begin{eqnarray*}
\frac{1}{\big(p^2-m_1^2\big)^n}&=&\theta(m_1^2-|p^2|)\frac{1}{\big[-(m_1^2+p_c^2)\big]^n}
\\&&+ \theta(|p^2|-m_1^2)\frac{1}{\big(p^2\big)^n},
\end{eqnarray*} we can obtain that
\begin{widetext}
\begin{eqnarray}
a(p^2) &=&
\theta(m_1^2-|p^2|)2a(m_1^2)\frac{|p^2|}{|p^2|+p^2_c}\left\{
\frac{3g^2(m_1^2)}{2\pi^2g^2(0)}\left(1-\frac{p^2}{m_1^2}\right)I_3(y_1,r)-1
-\frac{4\pi^2}{15}\frac{3{p^0}^2+{\vec{p}}^2}{p^2}\frac{T^2}{m_1^2+p_c^2}-\cdots
\right\}\nonumber \\
 &&+\theta(|p^2|-m_1^2)2a(m_1^2)\frac{m_1^2+p_c^2}{(|p|^2+p_c^2)\varepsilon(p^2)}\left\{
\frac{3g^2(p^2)}{2\pi^2g^2(0)}\left(1-\frac{m_1^2}{p^2}\right)I_3(y_1,r)\frac{m_1^2}{p^2}+1
+
\frac{4\pi^2}{15}\frac{3{p^0}^2+{\vec{p}}^2}{p^2}\frac{T^2}{p^2}+\cdots
\right\}.\nonumber \\
\end{eqnarray}
\end{widetext}
From Eq.(34) we can get the equation obeyed by $A(m_1^2)$
$$
A(m_1^2)=1-a(m_1^2)/A(m_1^2)
$$
which has the solution
\begin{equation}
A(m_1^2)=\frac{1}{2}\big\{1+[1-4a(m_1^2)]^{1/2}\big\}.
\end{equation}
When temperature $T$ is low enough so that $4a(m_1^2)<1$,
$A(m_1^2)$ will have the order of magnitude of one, and it is seen
from Eq.(38) that we always have $a(p^2)\ll 1$, whether $|p^2|\ll
m_1^2$ or $|p^2|\gg m_1^2$. Hence we can obtain
\begin{equation}
A(p^2)\simeq 1,\; \;\;\mbox{for}\;\;
\frac{T^2}{m_1^2+p_c^2}<\frac{9f(m_1^2)}{4g^2(0)}.
\end{equation}
When  temperature $T$ is neither very low nor very high so that
$4a(m_1^2)>1$ but $T^2/(m_1^2+p_c^2)<1$, the series expansion in
$T^2/(m_1^2+p_c^2)$ in Eq.(38) may still be used. In this case,
$A(m_1^2)$ becomes complex and we may write the module of
$a(p^2)/A(m_1^2)$ by
\begin{equation}
\left|\frac{a(p^2)}{A(m_1^2)}\right|=
\frac{|a(p^2)|}{|1+i\sqrt{4a(m_1^2)-1}|/2}
=\frac{|a(p^2)|}{a^{1/2}(m_1^2)}.
\end{equation}
Noting that the terms contained in the brace in Eq.(38) have the
order of magnitude of unity, the module may be approximately
estimated by
\begin{widetext}
\begin{equation}
 \frac{|a(p^2)|}{a^{1/2}(m_1^2)}
 \sim \theta(m_1^2-|p^2|)2a^{1/2}(m_1^2)\frac{|p^2|}{|p^2|+p^2_c}+
\theta(|p^2|-m_1^2)2a^{1/2}(m_1^2)\frac{m_1^2+p_c^2}{(|p|^2+p_c^2)}.
\end{equation}
\end{widetext}
In spite of $2\sqrt{a(m_1^2)}>1$, when $|p^2|>m_1^2$,
$|a(p^2)|/a^{1/2}(m_1^2)\ll 1$ is obvious if $p^2$ is large
enough. On the other hand, when $|p^2|<m_1^2$, we may make a
numerical estimation of  $|a(p^2)|/a^{1/2}(m_1^2)$ in practical
case of QCD. If the running gauge coupling $g^2(p^2)$ in QCD with
the infrared momentum cutoff $p_c^2$ is taken as \cite{kn:8}
$$
g^2(p^2)=2\pi^2A/\ln\frac{|p^2|+p_c^2}{\Lambda_{QCD}^2},
$$
where $A=24/(33-2N_f)$ and $\Lambda_{QCD}$ is the RG-invariant
mass scale parameter, then the quark mass function $m(p^2)$ (when
only dynamical quark mass is involved) can be expressed by
$$
m^2(p^2)=\Big(\frac{2\pi^2A}{3}\Big)^2\Big(\frac{\phi}{p^2+p_c^2}\Big)^2
\Big(\ln\frac{|p^2|+p_c^2}{\Lambda_{QCD}^2}\Big)^{A-2},
$$
where $\phi$ is the RG-invariant quark-antiquark condensates
\cite{kn:8}. Thus Eqs.(32) and (33) lead to
$$
m_1=\frac{2\pi^2A}{3}\frac{|\phi|}{m_1^2+p_c^2}
\Big(\ln\frac{m_1^2+p_c^2}{\Lambda_{QCD}^2}\Big)^{A/2-1}
$$
and
$$
f(m_1^2)=1+\frac{2m_1^2}{m_1^2+p_c^2}
\left[1+\frac{2-A}{2}\Big(\ln\frac{m_1^2+p_c^2}{\Lambda^2_{QCD}}\Big)^{-1}\right].
$$
It may be checked that, taking $p^2 = m_1^2/2$,  when $N_f = 3$
and $t_c\equiv \ln(p_c^2/\Lambda_{QCD}^2) = 0.1$,  for
$|\phi|/\Lambda^3_{QCD}$ = 0.1, we will have $m_1/\Lambda_{QCD}
\simeq 0.64$ and obtain $|a(p^2)|/a^{1/2}(m_1^2) = 0.5$ and 0.86
respectively for $T/(m_1^2+p_c^2) = 0.5$ and $T/p_c = 1$; for
$|\phi|/\Lambda^3_{QCD}$ = 0.5, we will have
$m_1/\Lambda_{QCD}\simeq 1.2$ and obtain $|a(p^2)|/a^{1/2}(m_1^2)
= 1.33$ and 1.75 respectively for $T/(m_1^2+p_c^2) = 0.5$ and
$T/p_c = 1$. These results show that when $|p^2| < m_1^2$,
$A(p^2)\simeq 1$ is not a good approximation. However, the
difference between $A(p^2)$ and 1 is not very large and it is easy
to verify that it will decrease as $T$ goes down and the infrared
momentum cutoff $p^2_c$ goes up. It is also seen that the value of
$m_1^2$ either lower or slightly higher than the infrared momentum
cutoff $p_c^2$, so the region in which $|p^2| < m_1^2$ is quite
limited, and it may be presumed that the deviation of $A(p^2)$ to
1 will cause only quite small effect to the total results. Hence
we can neglect such deviation and take $A(p^2)\simeq 1$ for all
the value of $p^2$ in the conditions
\begin{equation}
\frac{9f(m_1^2)}{4g^2(0)} < \frac{T^2}{m_1^2+p_c^2}<1.
\end{equation}
To sum up, we can assume that
\begin{equation}
A(p^2)\simeq 1, \;\mbox{for}\; 0\leq T^2 < m_1^2+p_c^2
\end{equation}
is a feasible approximation. When only dynamical quark mass is
involved, the minimal $m_1$ will be zero, and when a current quark
mass is also included, the minimal $m_1$ may still be quite small,
so the allowed highest temperature limited by Eq.(44) may be
reasonably supposed to be $T = p_c = e^{t_c/2}\Lambda_{QCD}=1.05
\Lambda_{QCD}$ (e.g. for $t_c=0.1$)  and  will increases as a
larger $p_c$ is taken. Hence, if the chiral phase transition
temperature is lower than $\Lambda_{QCD}$ \cite{kn:12,kn:19}, then
the above arguments indicate that within the whole range of
temperature related to chiral phase transition, $A(p^2)=1$ is a
feasible approximation and may be used in the discussion of chiral
phase transition problem in QCD based on Schwinger-Dyson approach.
This will certainly greatly simplify calculations in this
nonperturbative method.

\end{document}